\documentclass[conference]{IEEEtran}
\ifCLASSINFOpdf
\else
\fi
\usepackage{balance}
\usepackage{comment}
\usepackage{times,listings,amsfonts}
\usepackage{amsmath,xcolor}
\usepackage{graphicx}
\usepackage{pifont}
\usepackage{xspace}
\usepackage{url}
\usepackage{multirow}

\def\Quiver{Quiver\xspace} %Name of our system.
\begin{document}
\title{Quiver: Using Control Perturbations to Increase the Observability of Sensor Data in Smart Buildings}

\author{
\IEEEauthorblockN{Jason Koh, Bharathan Balaji, Vahideh Akhlaghi, Rajesh Gupta}
\IEEEauthorblockA{University of California, San Diego\\
 \{jbkoh, bbalaji, vakhlagh, gupta\}@eng.ucsd.edu}
 \and
 \IEEEauthorblockN{Yuvraj Agarwal}
\IEEEauthorblockA{Carnegie Mellon University\\
 yuvraj.agarwal@cs.cmu.edu}
 \and
}


\maketitle

\begin{abstract}
%Active control of CPSes can be leveraged as a tool to facilitate better modeling and innovative applications. We focus on buildings as an exemplary CPS and show that control perturbations can be leveraged to explore and model control systems with a much improved accuracy than data based analysis alone. We design Quiver, an experimental framework for actuation of building HVAC system that enables us to perturb the control system safely. Using Quiver, we demonstrate three applications using empirical experiments on a real commercial building -- co-location of data points, mapping of depency between points and learning the relationship between data points.

Modern buildings consist of hundreds of sensors and actuators for monitoring and operation of systems such as HVAC, light and security. To enable portable applications in next generation \emph{smart buildings}, we need models and standardized ontologies that represent these sensors across diverse types of buildings. Recent research has shown that extracting information such as sensor type with available metadata and timeseries data analysis is difficult due to heterogeneity of systems and lack of support for interoperability. We propose perturbations in the control system as a mechanism to increase the observability of building systems to extract contextual information and develop standardized models. We design Quiver, an experimental framework for actuation of building HVAC system that enables us to perturb the control system safely. Using Quiver, we demonstrate three applications using empirical experiments on a real commercial building -- co-location of data points, identification of point type and mapping of dependency between actuators. Our results show that we can co-locate data points in HVAC terminal units with 98.4 \% accuracy and 63 \% coverage. We can identify point types of the terminal units with 85.3 \% accuracy. Finally, we map the dependency links between actuators with an accuracy of 73.5 \%, with 8.1 \% and 18.4 \% false positives and false negatives respectively.
\end{abstract}


% \fi
\IEEEpeerreviewmaketitle

\section{Introduction}
\label{sec:intro}

%Cyber Physical Systems (CPS) enable interconnectivity across various subsystems in domains such as automobiles and smart grid, expanding the breadth of applications to include cross system data analysis and system level control optimizations~\cite{baheti2011cyber,rajkumar2010cyber}. Models and abstractions are a fundamental requirement for development of next generation applications that exploit system interconnectivity and torrents of data to create reliable, efficient and adaptive systems~\cite{lee2008cyber,derler2012modeling,balaji2015models}. Many CPS models and innovative applications have been demonstrated that exploit the domain knowledge and data available, but only limited control is exercised in practice~\cite{shi2011survey}.

Cyber-Physical Systems (CPS) represent continuing advance of instrumentation of existing systems (such as automobile~\cite{berger2014autonomous}) and networks (such as transportation~\cite{abid2011v}, energy grid~\cite{karnouskos2011cyber}) with a growing list of sensors and actuators. These sensing mechanisms (from tire-pressure meters to synchrophasers) provide a greater awareness of complete systems with increasingly finer resolutions of time and distance scales. Yet, deployment and maintenance of growing sensing infrastructure presents a significant challenge~\cite{derler2012modeling,karnouskos2011stuxnet,balaji2015models}, one that researchers have tried to address through analysis of the sensory data. Actuation also provides us with increased observational capabilities by actively modulating a system and observing responses. Active control has shown to be effective across disparate disciplines -- seismic structural design~\cite{symans1999semi}, aerodynamic stability analysis~\cite{epstein1989active} and fault tolerant control~\cite{blanke2006diagnosis}. However, control is seldom used across most systems and much of the literature is based on simulation studies. We propose that control perturbations be used across a variety of CPS applications such as information discovery, modeling, control optimization and privacy protection. 

In this paper, we explore use of perturbation control to address sensor data management problem. Our thesis is that a carefully selected and controlled modulation of control algorithms can be used to discover sensor conditions and reduce sensor maintenance work required for creation of CPS applications. We focus on buildings as a driver application to test this hypothesis. Buildings consist of hundreds to thousands of sensors and actuators that are used for management of air conditioning, lighting, fire safety and water. Prior work has addressed many challenges to enable development of smart building applications -- standardized API for information access~\cite{dawson2010smap}, data management~\cite{arjunan2012sensoract,krioukov2012building}, and semantic ontology~\cite{haystack}. These solutions have led to innovative applications such as occupancy based control~\cite{balaji2013sentinel,gao2013spot}, human in the loop control~\cite{erickson2012thermovote,krioukov2011living} and energy disaggregation~\cite{jiang2009experiences}. A major challenge in adoption of these applications is that they are not portable across different building systems due to lack of standardized ontologies~\cite{balaji2015zodiac, bhattacharya2015metadata}.

Recent works have focused on standardizing building information using existing sensor metadata and available timeseries data to enable portable applications~\cite{balaji2015zodiac,bhattacharya2015metadata,gao2015metadata,hong2015adapter}. These works illustrate that modern buildings consist of a wide variety of sensors and actuators with varying naming conventions across vendors. Oftentimes the metadata available is not understandable to a non-expert and sometimes no metadata is available to understand the data context. Even when researchers used the available metadata, timeseries data and applied state-of-the-art algorithms to identify the \emph{type} of the sensor, the overall accuracy was low~\cite{gao2015metadata,hong2015adapter} or required significant manual effort~\cite{balaji2015zodiac,bhattacharya2015metadata}. Identifying sensor type, however, is just one step towards creation of standardized models that can be used by portable applications. Other pertinent problems include determining sensor location, relationship among sensors/actuators and models which capture the behavior of the system. %\BB{NIST and energy consumption references?}

Active control is a promising approach to address the lack of information available, as carefully designed control perturbations can reveal insights into system behavior that is not observed in regular operation. Recently, Pritoni et al.~\cite{pritoni2015discovering} showed that the mapping between the Air Handler Units (AHU) and the corresponding terminal units in the building Heating, Ventilation and Air Conditioning (HVAC) system can be inferred with 79\% accuracy with control perturbations compared to 32\% accuracy with data analysis alone. Control perturbations have also been studied for Fault Detection and Diagnosis (FDD)~\cite{weimer2012active,padilla2015combined} and fault tolerant control~\cite{fernandez2009self,padilla2015model} in HVAC systems as it eliminates mundane manual testing and fixes some classes of faults automatically. 

We expand on these ideas and show that active control mechanisms can be used as an integral part of a data model. Control based interventions are not used in practice because of equipment and safety issues. We empirically explore control in a real building HVAC system. We build \Quiver, a control framework that allows us to do control experiments \emph{safely} on the HVAC system by constraining control input that satisfies criteria such as range of values, frequency of actuation and dependency between actuators. We deploy \Quiver in our building testbed and use it to demonstrate three example applications that exploit control perturbations. First, we show that perturbations can be used to identify co-located sensors which has been shown to be difficult with data alone~\cite{hong2013towards}. We co-locate data points in HVAC terminal units with 98.4\% accuracy and 63\% coverage. Second, we identify the point type of terminal units given ground truth point types of one terminal unit using transfer learning classification. We identify point types with 85.3\% accuracy across 8 zones. Third, we map the dependency between sensors and actuators in the control system using control perturbations and probabilistic analysis. We identify dependency links between actuators with 73.5 \% accuracy, with 8.1\% false positives and 18.4\% false negatives across 5 zones.%Third, we model the relationship between sensors and actuators using perturbations, and demonstrate that these models are better than those created with data analysis alone. \BB{Change last paragraph based on results/contributions}

%\BB{Results line}

\section{Our Building Testbed}
%We focus on HVAC systems in buildings where hundreds of sensors and actuators are networked together for operation and maintenance. 

%Sensors include room temperature, air flow, water flow, fan speed, static pressure, etc. Actuators

Modern buildings consist of hundreds of networked sensors and actuators for operation and maintenance of various systems. These systems are typically overseen with Building Management System (BMS) which helps configure, monitor, analyze and maintain various systems. The sensors, actuators and the configuration parameters in the BMS are together referred to as \emph{points}. We focus on building HVAC systems where BMSes are most commonly used.

%Problems with BMS
%Traditional BMSes are vertically integrated systems and do not support third party applications or interoperability across vendors. This has led to vendor specific and institution specific naming conventions~\cite{rochesternaming,dodnaming} causing heterogeneity across buildings even for similar types of equipment~\cite{balaji2015zodiac,hong2015adapter}. To enable portable building applications, we not only need a common naming standard for different types of points, but also require a semantic model of the system which include contextual data such as location of points, role of each point in the building control system and a behavioral model of the system itself. We examine a five floor building in our university, and illustrate the challenges in inferring such contextual information from a real BMS based on our past research experiences and discussions with university Facilities Management. %We then show that control perturbations can be exploited to ease the analysis.

%\subsection{Our Building Testbed}

\begin{figure}[Ht]
\includegraphics[width=\linewidth]{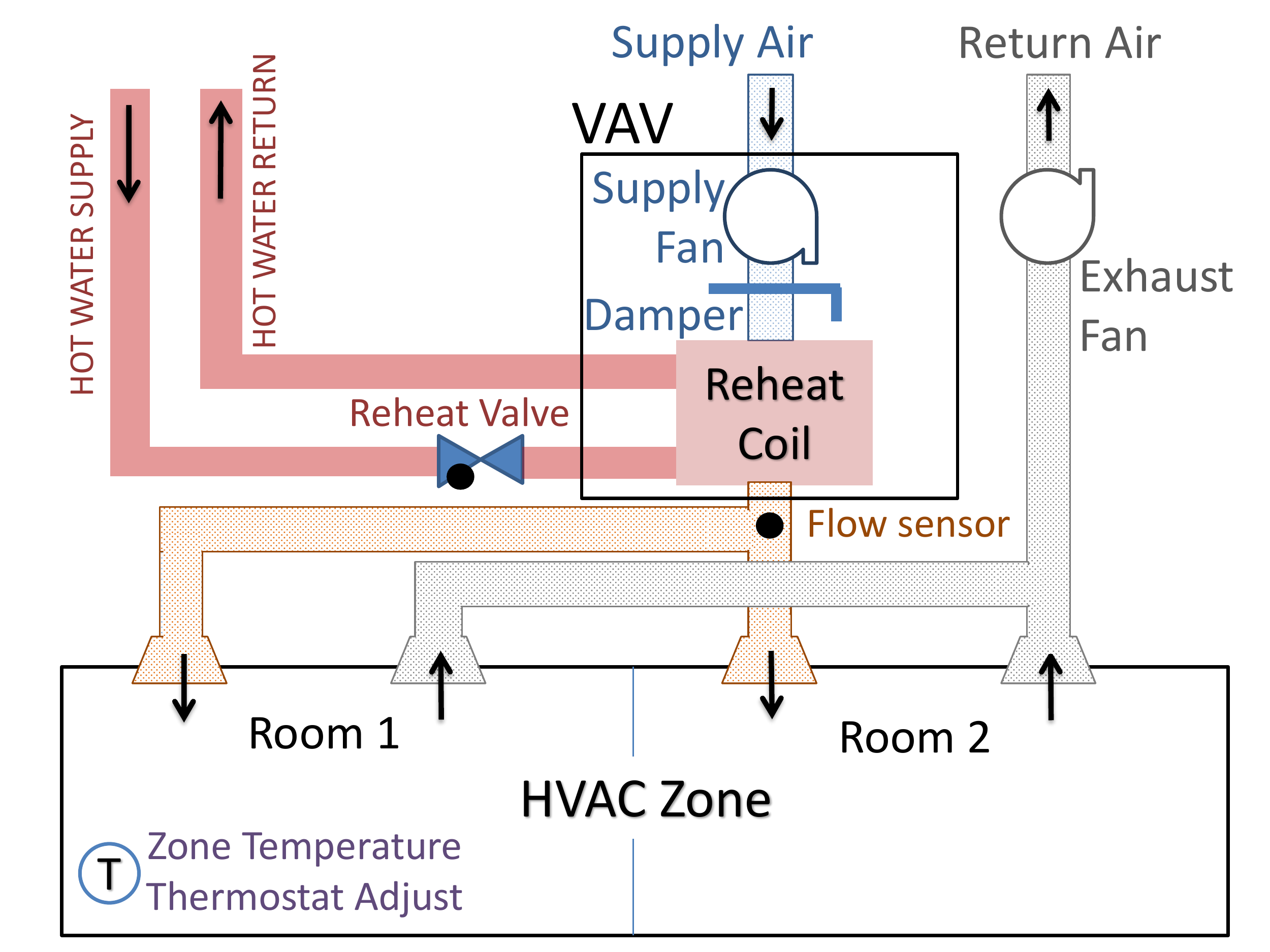}
%\vspace{-5mm}
\caption{Sensors and actuators in a Variable Air Volume (VAV) unit that provides local control of temperature in the HVAC system~\cite{balaji2013zonepac}.}
\label{fig:vav}
%\vspace{-5mm}
\end{figure}

Our testbed is a 150,000 sq ft building, constructed in 2004 and consists of a few thousand occupants and 466 rooms. The HVAC system consist of an Air Handler Unit (AHU) that supplies cool air to the building via ductwork using chilled water supplied by a central plant. A heat exchanger supplies hot water to the rest of the building using hot water supplied by central plant. The cool air and hot water are used by local terminal units called Variable Air Volume (VAV) boxes to regulate the temperature of rooms. The area serviced by the VAV box is referred to as a \emph{thermal zone}, which consist of a large room or multiple small rooms in our building. Figure \ref{fig:vav} shows a schematic of the VAV box with the sensors and actuators installed for its operation.

%The \texttt{Zone Temperature} sensor provides feedback to the VAV on the current status of the thermal zone. The VAV regulates the 
VAVs have been commonplace since 1990s~\cite{hydeman2003advanced}, and their basic working is well understood. The VAV regulates the amount of cool air provided using a damper, and if the zone needs to be heated, it regulates the hot water in the heating coil using a valve. The temperature sensor in the thermal zone provides feedback on how much cooling or heating is required. However, in the real VAV box, there are over 200 points that govern its working~\cite{vav2003points}. The essential sensors include: \textit{Zone Temperature, Supply Air Flow, Reheat Valve Position} and \textit{Damper Position}; and the actuator points include: \textit{Reheat Valve Command, Thermostat Slider Adjust} and \textit{Damper Command}. These actuators are controlled using many configuration points such as \textit{Temperature Setpoint, Occupied Command, Air Flow Setpoint}, etc. These configuration points account for majority of the points, and include nuanced parameters that ensure minimum airflow, set the PID loop settings, etc. 

Not all of these 200 points are reported to the BMS, and only the essential sensors and control points are exposed to limit resource usage and information overload for building managers. In our building testbed, 14 to 17 points are reported to the BMS for each VAV box. The points exposed to BMS changes depending on the vendor, type of VAV and the installation version used by the vendor. Even though the same model of VAV is used across all zones in our building, there are minor variations due to configuration changes, presence of supply/exhaust fans or lack of heating. 

%In our campus, there are over 55 buildings managed by the same vendor, and some of the older buildings do not have essential sensors such as \textit{Supply Air Temperature} or \textit{Supply Air Flow}, while others have detailed PID loop setting such as \textit{Heating Integration Time}. The BMS does not have information on the type of VAV box, or how these points relate to each other, and a skilled operator is expected to know the underlying context~\cite{teraoka2014buildingsherlock}. In our building, 

%Such heterogeneity in the BMS makes it challenging to develop standardized models that can be used for portable building applications. Simplistic models that only capture high-level working of the control system fail to exploit the rich information and control available in practice. We need methods to develop system models that encapsulate the functionality of the BMS points, and we propose that control perturbations can be a useful tool to capture the system behavior across different operating points.

\subsection{Data Collection and Control}

\begin{figure}[Ht]
\includegraphics[width=\linewidth]{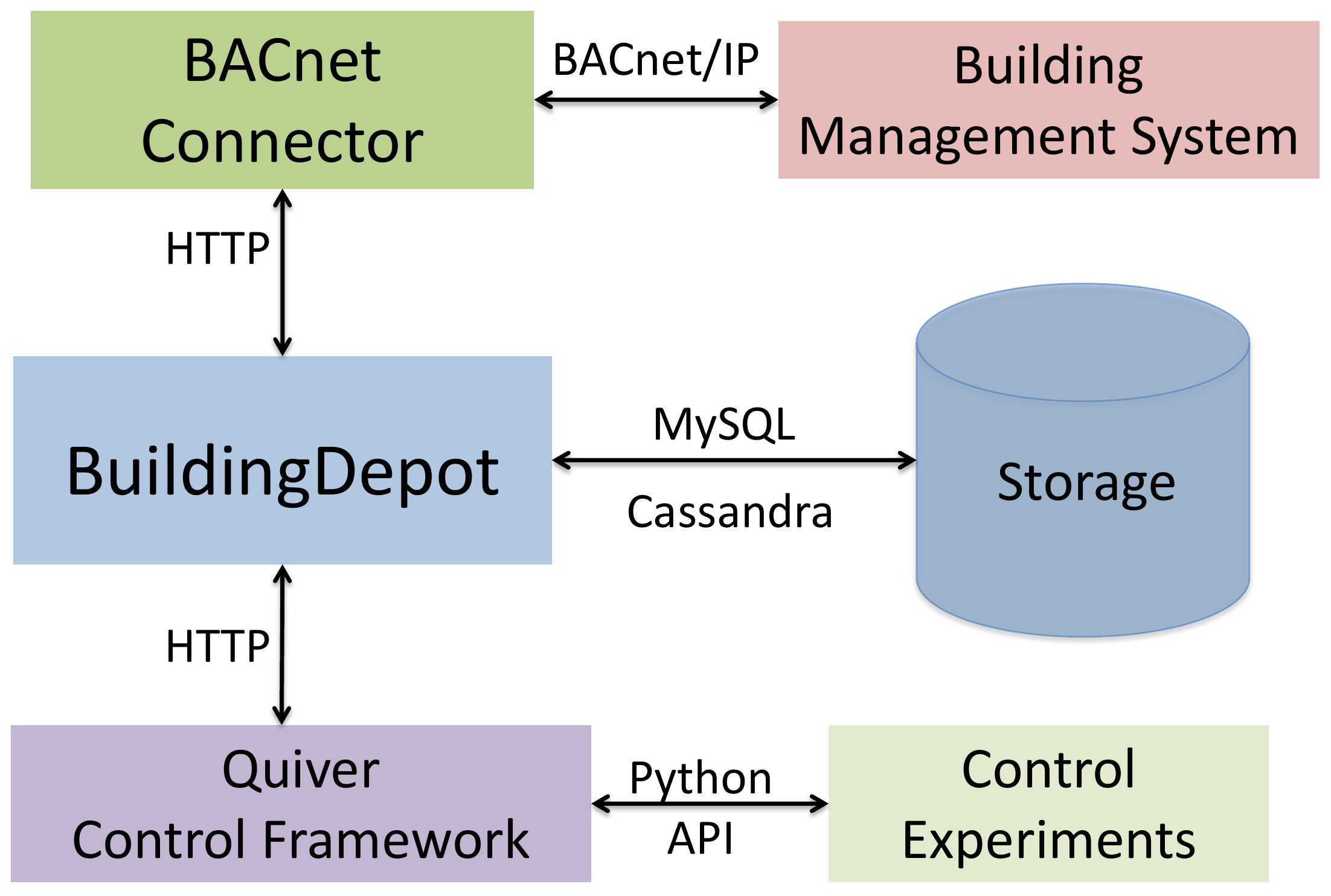}
%\vspace{-5mm}
\caption{System architecture of Quiver. Data collection and control is done via BACnet protocol using BuildingDepot web service~\cite{weng2013buildingdepot}. Quiver ensures that the control sequences of our experiments are safe and rolls back the system to its original behavior in case of failure.}
\label{fig:sys_arch}
%\vspace{-5mm}
\end{figure}

The points in our building communicate with the BMS using BACnet~\cite{bushby1997bacnet}, a standard building network protocol. We connect our server to this network to collect data and control the points in our building. We use BuildingDepot~\cite{weng2013buildingdepot}, an open source RESTful web service based building datastore to manage the points in the building, provide appropriate permissions to developers, and search using a tagging mechanism. Our control framework Quiver works on top of BuildingDepot to manage control inputs from our experiments. Figure \ref{fig:sys_arch} depicts the system architecture of our deployment.

BACnet is a well developed protocol with which developers can not only read and write points, but also schedule hourly control, mark holidays on a calendar, and even manage programs running in the embedded VAV controller. For simplicity, we only focus on read and write points, i.e., in BACnet terminology \emph{Input, Output} and \emph{Value} points. These points can have floating point, binary or multi-state values, and in BACnet a floating point that can be written to is referred to \emph{Analog Output}. Each of these \emph{Output} points have an associated priority array. The default operation is performed at the lowest priority and the highest levels are reserved for emergency operations such as fire safety. Once a higher level priority is written to, the lower levels are ignored. An explicit write with value ``0'' needs to be written to the higher level priority in order to relinquish control to the lower levels. 

The university Facilities Management provides us with a fixed priority level in this priority array for our control experiments. We need to relinquish control back to the default priority level after our control experiments to ensure that our interference does not affect the regular operation of the HVAC system. Quiver ensures that all the points are relinquished after an experiment.

\subsection{Points in Variable Air Volume Box}

%We have been working with our building testbed for the past six years and we build our Quiver control framework based on vendor provided manuals, discussion with building managers, empirical control experiments and our past experiences. We focus on the VAV box in the building HVAC system to demonstrate the feasibility of control experiments, and will pursue control of other equipment such as AHU in future work. 

\begin{figure*}[Ht]
\includegraphics[width=\linewidth]{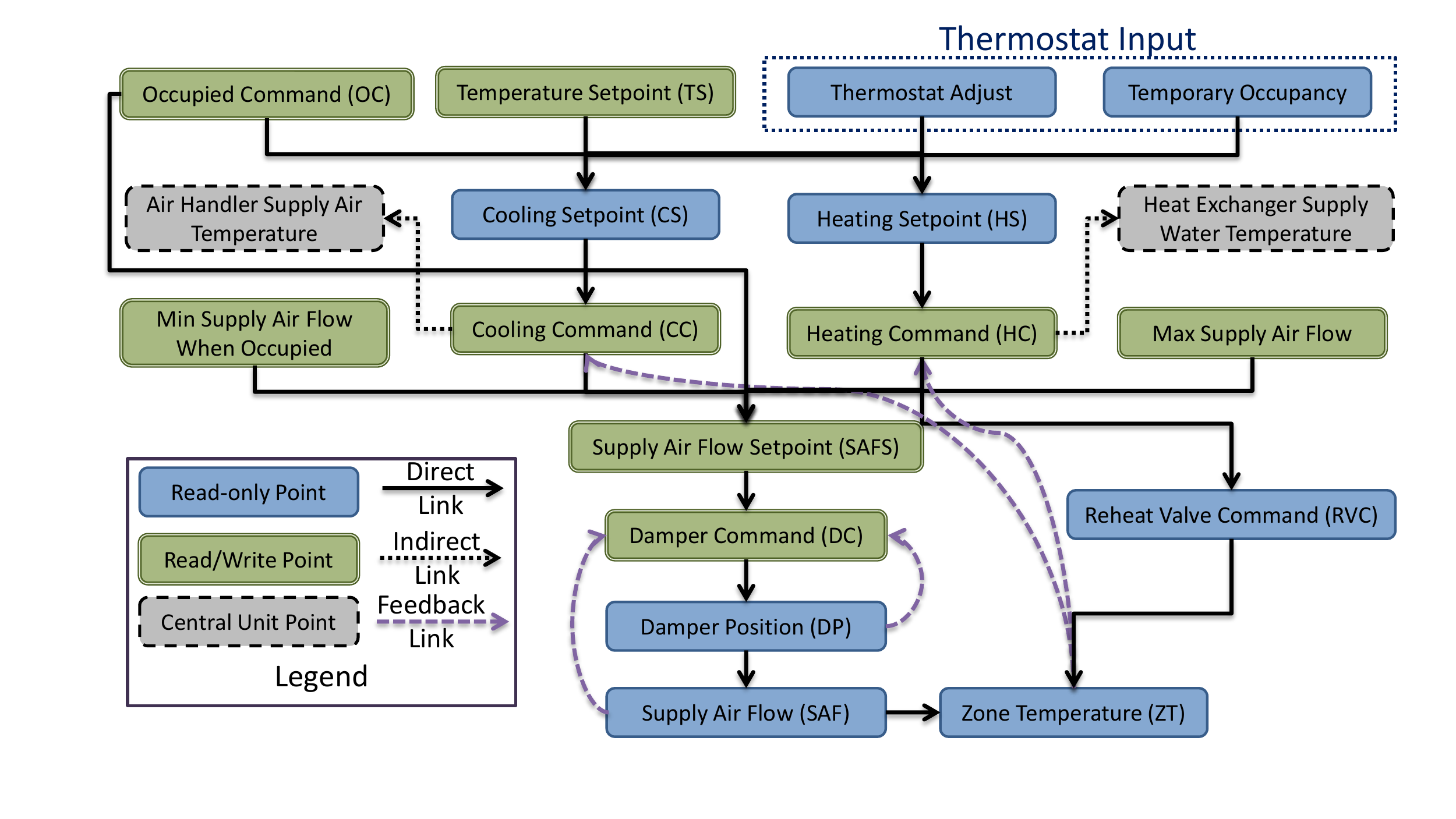}
%\vspace{-5mm}
\caption{BMS points associated with VAV in our building testbed. The dependency between the points as shown by arrows is mapped based on domain knowledge. Read-only points are either sensors or configuration points which cannot be changed. Read/write points can be changed via BACnet.}
\label{fig:dependency_vav}
%\vspace{-5mm}
\end{figure*}

Figure \ref{fig:dependency_vav} shows the points associated with VAV in our building BMS and how these points relate to each other. At the top of the figure, we have the zone \emph{Temperature Setpoint} and \emph{Occupied Command}, which in combination with thermostat input determine the temperature guardband within which the VAV is trying to keep the zone temperature. The temperature guardband is indicated by \emph{Heating} and \emph{Cooling Setpoints}, which represent the lower and upper bounds of temperature respectively. There are three occupancy modes: \emph{Occupied}, \emph{Standby} and \emph{Unoccupied} during which the temperature bands are $4^oF$, $8^oF$ and $12^oF$ respectively. During the \emph{Occupied} mode, minimum amount of airflow is maintained to ensure indoor air quality. The \emph{Thermostat Adjust} allows changing the temperature setting by $\pm1^oF$, and the \emph{Temporary Occupancy} maps to a button on the thermostat which when pressed puts the zone to \emph{Occupied} mode for two hours during nights/weekends. %Note that we do not have explicit humidity control in our building testbed.

The \emph{Heating} and \emph{Cooling Setpoints} determine the behavior of the VAV control system with the measured \emph{Zone Temperature} completing the feedback loop. These three points determine the \emph{Cooling} and \emph{Heating Command} of the thermal zone. The \emph{Cooling Command} determines the amount of cool air required for the zone and determines an appriopriate \emph{Supply Air Flow Setpoint} that is between the designed minimum and maximum supply air flow. When the \emph{Cooling Command} is high ($\sim100\%$), feedback is sent to the AHU to decrease the supply air temperature to meet the cooling needs of the thermal zone. The \emph{Heating Command} determines the amount of reheat required by controlling the \emph{Reheat Valve Command}. During heating, the airflow is set to the minimum to reduce chilled airflow from AHU, and this airflow is increased when high \emph{Heating Command} ($\sim100\%$) fails to heat up the thermal zone sufficiently. A high \emph{Heating Command} also sends a signal to the heat exchanger to increase the supply water temperature. Note that only one of \emph{Heating} or \emph{Cooling Commands} can be \textgreater 0\% at a time.

The \emph{Supply Air Flow Setpoint} determined by the cooling/heating requirements in turn determines the \emph{Damper Command} which is the amount of damper actuation required to match the setpoint to the measured \emph{Supply Air Flow}. The \emph{Damper Position} sensor also provides feedback to set the appropriate \emph{Damper Command}. There is a separate PID loop associated with setting each of \emph{Heating Command, Cooling Command, Supply Air Flow Setpoint} and \emph{Damper Command}, and there are PID parameters such as those that govern proportional gain and integration time, but these are hidden from the BMS. %These points are hidden from the BMS and can be only set by the building managers using a vendor provided proprietary software. In addition to these points, other types of VAVs can have points related to control of exhaust flow, supply/exhaust fan, fume hood flow and humidity. We ignore these as they are not present in our building.
\section{Quiver: Control Framework}
\label{sec:framework}

Quiver\footnote{The basic control framework will be presented as a poster abstract (not peer reviewed) in an upcoming conference.} is built upon our building testbed. In this section, we show the utility of Quiver on the VAV box to demonstrate our control perturbation applications. 

%Our control framework needs to understand this detailed working of the VAV in order to provide fine-grained control to developers. We focus on providing a safe control environment provided the control is performed using a single Quiver server, and we discuss how to overcome this limitation in future work. The developer can run multiple experiments at a time, and Quiver analyzes the control sequence for data errors and dependency checks. 

Given that exercising control over a building control system can lead to unintended, and potentially dangerous consequences, we ensure safety for the control perturbations we perform by: %We ensure safety of the control perturbation through following means: 
(a) global time synchronization across the computational servers running different parts of Quiver, (b) identification of a range of `safe' values for control through metadata and data history, (c) dependency checks based on domain knowledge, and (d) status tracking to ensure we relinquish control at the end of our experiment and are able to rollback the changes made in case of an application crash. We have designed these safeguards based on five years of control experience with BMSes and discussions with our university Facilities Management.
%management and rollback: we ensure that all setpoints are explicitly set or reset and are guaranteed to rollback in case of exception conditions regardless of the source of exceptions. %<Now describe each of these. Is there a notion of setpoint commit state so taht partially completed operations do not lead to setpoint commitments?>i
%\BB{Needs revision}

We perform basic checks such as checking the range of values exercised (i.e. min/max), type of data for each input based on metadata available in BACnet as well as self-imposed limits. For instance, we force the \emph{Temperature Setpoint} to be between $62^oF - 78^oF$. Each point marked as read/write in Figure \ref{fig:dependency_vav} can be written to using Quiver. We synchronize time across all our servers -- BACnet Connector, BuildingDepot and Quiver -- using our university Network Time Protocol server. The time synchronization across these servers is necessary to ensure correct sequence of operations and proper data analysis. 

A local database keeps track of all the read/write points in the testbed building. For example, we keep a track of whether the point is part of a current experiment, the last value that was written to the point, the reset (default) value of the point, the timestamp of the last write to that point, and the thermal zone that the point belongs to. This database is used to keep track of the status of the experiment as well as allow developers to query it through APIs. The database is also used to ensure that all points are reset to their default values at the end of an experiment and, if an experiment crashes due to an error condition, a rollback is performed to restore the points to a safe (default) state.   

Quiver limits the frequency of control of a single point, or dependent points, to allow time for HVAC control changes to take effect and to avoid problems such as damper oscillations which reduces equipment life. Currently, we have set the minimum time between consecutive writes to 10 minutes. Before writing to a BACnet point, Quiver checks for dependency between the point and other points which have been written to by referring the local database. We \emph{conservatively} assume that all the control points in a particular VAV are related to each other, and thus, only one control input can be written to a VAV every 10 minutes. Note that there are 237 thermal zones in the building, and each of these zones can be controlled in parallel. We set a minimum delay of 10 seconds between each write so that the BMS does not get overloaded. After each write, Quiver ensures that the BMS has accepted the input and throws an exception in case the write is rejected after retrying two times. 

Upon further experimentation, we found that the VAV embedded controller has some built-in safety features that also limit the amount of control available. For example, the VAV does not allow the \emph{Supply Air Flow Setpoint} to be set beyond the minimum and maximum values. It also does not allow heating when the \emph{Zone Temperature} exceeds the \emph{Heating Setpoint} and disallows cooling when the temperature is lower than the \emph{Heating Setpoint}. The \emph{Damper Command} represents the change in \emph{Damper Position} required, and resets itself after the damper is actuated appropriately. This behavior is different from the rest of the actuator points such \emph{Supply Air Flow Setpoint}, which match the sensor value as much as possible. We incorporate these constraints in our experiments, and integrate them in the next version of the control framework. %\YA{next version? are they not there now?} 

%\YA{I still feel that you could make this section a bit more exciting by discussing how you came up with the safeguards that you did, whether there are some important ones missing, etc. right now it seems to be just stating what we did, rather than why we did it or why it was needed?}. 

\section{Quiver: Control Experiments}

We use Quiver to learn more information about the sensor and actuator points inside a building using control perturbations. We define control perturbation as any changes made to actuators that deviates from typical HVAC operation. We confine all of our control experiments to nights/weekends, or in unoccupied zones only, to alleviate any effects on occupant comfort. We focus our control experiments towards addressing three important smart building applications:

\begin{itemize}
	\item Identifying points which are co-located with a VAV box.
	\item Identifying point types within a VAV given ground truth point types of one VAV unit.
	\item Mapping the dependency between VAV actuator points.
%	\item Modeling the relationship between points using data driven regression.
\end{itemize}

All of our data analysis is implemented using Python Scikit Learn library~\cite{scikit-learn}.

\begin{figure}[Ht]
\includegraphics[width=\linewidth]{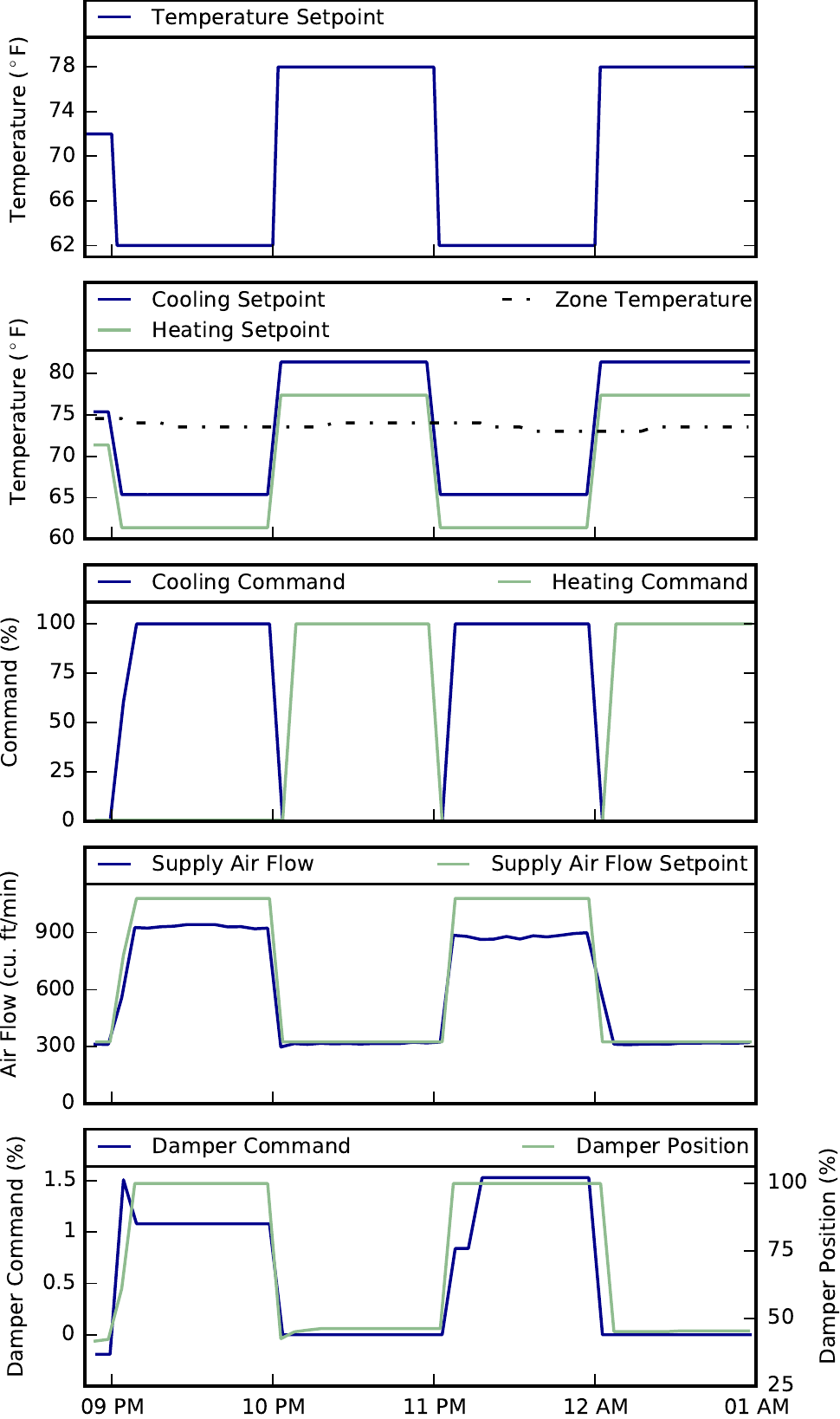}
%\vspace{-5mm}
\caption{A sample of co-location experiment, where the \emph{Temperature Setpoint} is oscillating between $62^oF$ and $78^oF$ for four hours at night (the top graph). The VAV points which react to these changes (remaining four graphs) can be co-located by using temporal data analysis of this controlled period.}
\label{fig:sample_control}
%\vspace{-5mm}
\end{figure}

\subsection{Using Control Perturbation to Determine Location}
\label{sec:colocation}

The location of sensor and actuator points is not readily available in the BMS for older buildings. In buildings where location information is available, it is usually inconsistent due to errors in manual labeling  process~\cite{bhattacharya2015metadata,gao2015metadata}. It is also difficult to co-locate points using historical data alone as many VAVs function in a similar manner, and the variation of data is not enough to distinguish them apart~\cite{hong2013towards}. Control perturbations can be used to force the control system to unusual operating points, and co-located points that respond to this perturbation can be clustered together by data analysis. %Thus, control perturbations can be used to create this missing variation in data to identify co-located points. \YA{Not sure I understand what missing variations there are?} 

We assume that we already know the type of points in the building which can be obtained using recently proposed methods~\cite{bhattacharya2015metadata,hong2015adapter}, but do not know if they are co-located or how these point relate to each other or affect the control system. We do not use the location information already integrated into Quiver for these experiments. We perturb the actuator point identified as the \emph{Temperature Setpoint} (TS) of a randomly chosen zone, and identify the corresponding co-located points using the temporal data features of other points. Towards the end of this section, we discuss how we can relax the assumption of knowing the point type apriori.

Figure \ref{fig:sample_control} shows an example control sequence, where we change TS 4 times across 4 hours from low ($62^oF$) and high ($78^oF$), and the corresponding VAV points in the same zone that react to its changes. We chose such an oscillation of TS as it deviates substantially from normal operation and we can easily distinguish the controlled zone from the rest of the zones under normal operation. This control sequence was chosen empirically, and we show that even such simple control sequences can be effective for co-location of points. However, as we show with our experiments, the effect of control sequences do affect the quality of results. We do not focus on design of generic control sequences in this paper. %\YA{I don't understand Figure 4. Which graph shows the 62-78 perturbation and 4 times?} 

%This signal is differentiable among users' normal behavior in terms of range and frequency. Our analysis on users' behavior in this building shows that a thermal zone's configurations (TS, OC, thermostat adjust, and temporary Occupancy) change 0.112 times per day in average with standard deviation 0.235. Also, TS can be changed by thermostats, which only changes $2^oF$ in most cases. \JK{Should we talk about thermostats like Genie here rather than just thermostats?} Those observations indicate that frequent and large oscillation of actuator and sensor points will be distinguishable from the other uncontrolled points. 

We extract basic features such as range (max - min), mean and standard deviation from the observed timeseries data. We also extract the Dynamic Time Warping (DTW) distance~\cite{berndt1994using} between the applied TS signal and the point under consideration. DTW compensates for the time delay in the reaction and change in sensor values due to a control action and quantifies the difference between the shape of the signals. In addition, we exploit our pulse control and analyze the Fast Fourier Transform (FFT) after normalizing the data and use the Euclidean distance between the FFT of the point signal and FFT of the TS signal. We refer to this feature as ``L2 norm of FFT'' or ``LFT''. We ignore frequencies beyond 0.0005 Hz, i.e. a period of 30 minutes, because we only focus on changes caused by our low frequency control sequence. 

We extract these features for all the VAV points in the building and identify the outlier points. In principle, the point which deviates the most from regular control operations would be co-located with our TS point with high probability.%\YA{mHz? I don't get the last 2 sentences.} %our concern is differentiating our signals from user behaviors, which hardly happen more frequently than 30 minutes periodically as we observed users' behavior.

%(((Followings are about 8 pulses))) %Figure \ref{fig:sample_control} shows an example control sequence, where we change TS 8 times across 12 hours and the corresponding points in that zone react to its changes. We oscillate the TS between low ($64^oF$) and high ($76^oF$) value to maximize the variation in the data. We extract basic features such as range (max - min), mean and standard deviation. We also extract the Dynamic Time Warping (DTW) distance~\cite{berndt1994using} between the applied TS signal and the point under consideration. DTW compensates for the time delay in the reaction and quantifies the difference between the shape of the signals. In addition, we exploit our pulse control and analyze the frequency spectrum using FFT after normalizing the data. We ignore frequencies beyond 10Hz, and use the Euclidean distance of the FFT between the point signal and the TS signal. We refer to this feature as ``L2 norm of FFT'' or ``LFT''.

\begin{figure}[Ht]
\includegraphics[width=\linewidth]{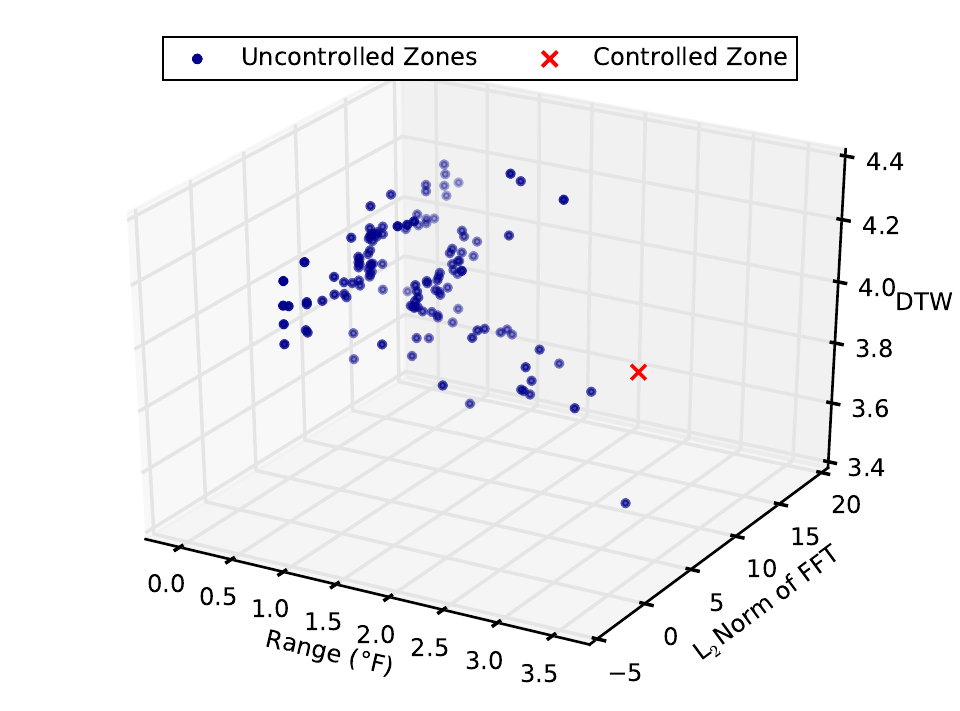}
%\vspace{-5mm}
\caption{Co-location of Zone Temperature by perturbing the Temperature Setpoint. 2 changes of Temperature Setpoint is applied across 2 hours in the controlled zone.}
\label{fig:colocation_1_pulses}
%\vspace{-5mm}
\end{figure}

\begin{figure}[Ht]
\includegraphics[width=\linewidth]{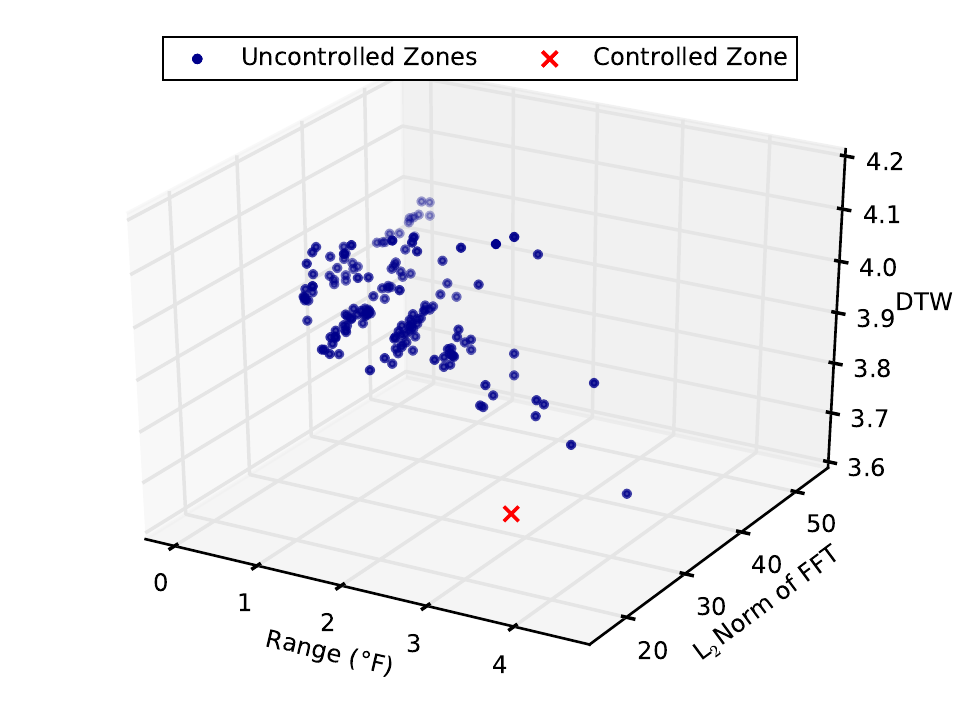}
%\vspace{-5mm}
\caption{Co-location of Zone Temperature by perturbing the Temperature Setpoint. 4 pulses of Temperature Setpoint is applied across 4 hours in the controlled zone.}
\label{fig:colocation_2_pulses}
%\vspace{-5mm}
\end{figure}

Figure \ref{fig:colocation_1_pulses} shows the distribution of all the \emph{Zone Temperature} (ZT) points in our building across three features -- DTW, LFT and range -- with a control sequence of two changes to TS. The zone under control is marked in red, and as observed, the red point is far away from most of the points from the other zones in the building. However, there are still a few points which are also differ significantly from most zones and it is difficult to distinguish the red point from those outliers. When we examine the data from the experiment where we made 4 changes to TS, the corresponding to changes to the ZT in the same zone resulted in much higher variation from those in uncontrolled zones. This is captured by our features as shown in Figure \ref{fig:colocation_2_pulses}. Hence, with the help of a well designed control perturbation it is possible to mold the behavior of the control system for end use applications. %\YA{I have read this para twice and just don't get it...} 

\begin{figure}[Ht]
\includegraphics[width=\linewidth]{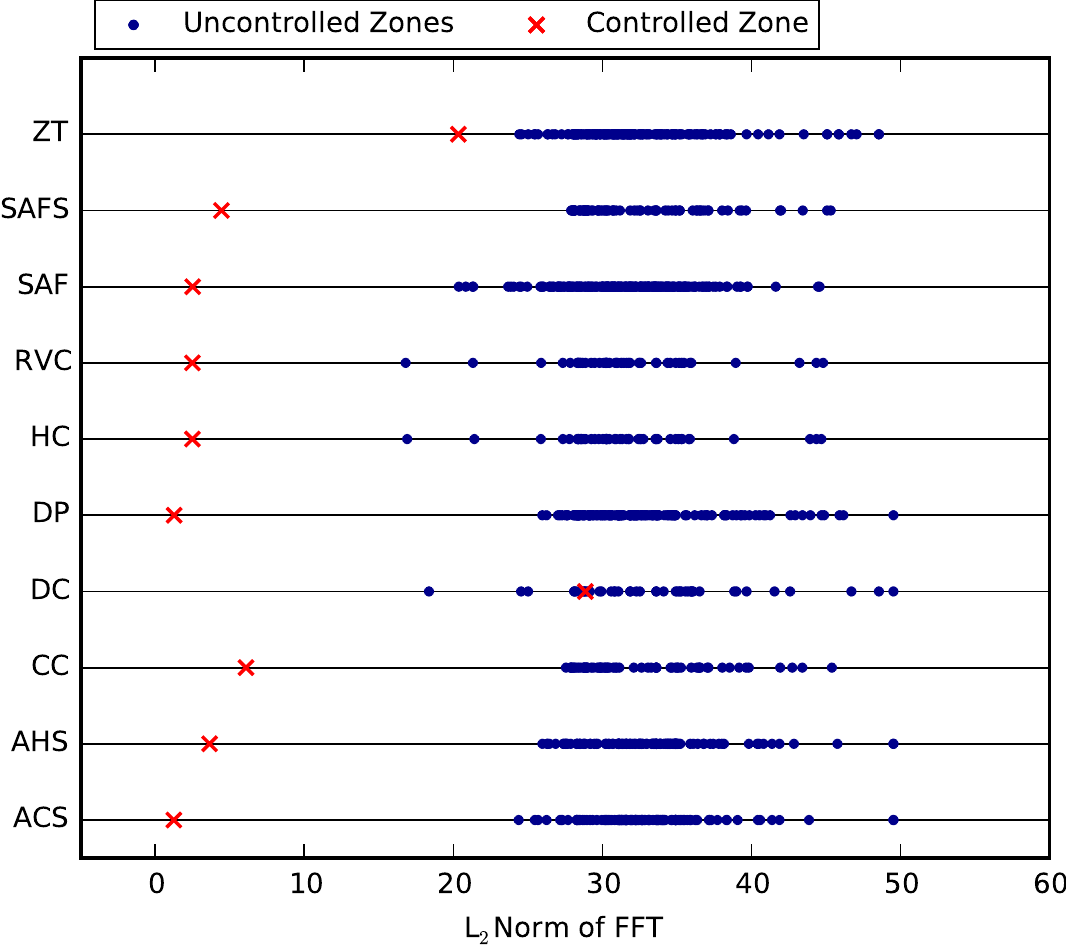}
%\vspace{-5mm}
\caption{Comparison of the L2 norm between FFT of VAV points and the FFT of the controlled \emph{Temperature Setpoint}. The points corresponding to the controlled VAV have a much lower L2 norm compared to regular zones for eight point types, but fails to capture the difference in \emph{Damper Command}.}
\label{fig:colocation_zone}
%\vspace{-5mm}
\end{figure}

We analyze the data for other point types to check if we can co-locate the zonal points successfully. In practice, we find that LFT feature alone is sufficient to distinguish the controlled zone points from the rest. Figure \ref{fig:colocation_zone} shows compares the LFT of the controlled zone with other zones for point types: \emph{Zone Temperature} (ZT), \emph{Supply Air Flow Setpoint} (SAFS), \emph{Supply Air Flow} (SAF), \emph{Reheat Valve Command} (RVC), \emph{Heating Command} (HC), \emph{Damper Position} (DP), \emph{Cooling Command} (CC), \emph{Heating Setpoint} (HS) and \emph{Cooling Setpoint} (CS). We performed this control experiment on eight zones in our building, and we co-located the listed points with 98.6\% accuracy. We only failed to identify the correct \emph{Damper Position} point for one of the zones, leading to a drop in accuracy.

The \emph{Damper Command} (DC) is a differential actuator that sets the change that needs to be made to the damper. There are several VAVs in the building which constantly change their DC for minor variation in the air flow, and the features we extracted -- DTW, FFT, mean, variance, number of changes -- failed to differentiate the DC of the zone under control from the rest (Figure \ref{fig:colocation_zone}). More sophisticated data analysis or perturbance signals are required for co-location of DC points. We could only co-locate DC points in two of the eight zones with our current method. %We pursue different perturbation signatures for these type of points in future work.

Another issue with these control experiments is that we can only co-locate those points which react to changes in TS (see Figure \ref{fig:dependency_vav}). Points such as \emph{Occupied Command} (OC) and \emph{Thermostat Adjust}, which are external inputs to the VAV control system, cannot be co-located. To remedy this, we perform a second set of experiments which oscillates the OC similar to our TS control perturbations. We successfully co-located OC and points such as \emph{Cooling Setpoint} and \emph{Supply Air Flow} that have been already co-located with their corresponding TS point. Thus, all of these points can be marked as being co-located in the same VAV. We performed the \emph{Occupied Command} oscillation experiments across four zones with 100\% success rate in their co-location results. We could not perform similar experiments on the thermostat points (\emph{Thermostat Adjust} and \emph{Temporary Occupancy}). They cannot be controlled by our platform as thermostats produce their data contiguously, and we acknowledge this is a limitation of our proposed method. %\YA{what's a thermostat point that cannot be controlled by Quiver?}. 

We repeated our TS control experiments on a hot day, and found that the same control perturbations cannot co-locate heating related points -- \emph{Heating Command} (HC), \emph{Reheat Valve Command} (RVC) -- as they are not triggered sufficiently due to hot outside weather. The zone cannot be cooled down enough to activate heating (HC, RVC) when its TS is changed to the high value. We need to change our control perturbations to excite these points specifically. Thus, the perturbation signature needs to be sensitive to external conditions and confounding factors. %We changed our control signature .. \BB{Add text after performing this experiment}

Note that in Figure \ref{fig:colocation_zone}, the LFT feature of the controlled zone for most point types except ZT and DC differ significantly from the rest of the points. Controlled zone's LFT feature of ZT is not distinguishable from other types' LFT feature of the other zones though it can be co-located within the same type. Contrary to DC's different operational behavior, ZT's signal response is slower than the other types due to heat capacity of zones. If we assume that we do not know the point type, then the points except ZT and DC can be obtained as outliers from the points belonging to normal zones. However, we would need to design appropriate threshold or clustering technique to identify the outlier points correctly.

%Note that in Figure \ref{fig:colocation_zone}, the LFT feature of the controlled zone for all point types differ significantly from the rest of the points. If we assume that we do not know the point type, then these points can be obtained as outliers from the points belonging to normal zones. However, we would need to design appropriate threshold or clustering technique to identify the outlier points correctly. \JK{Current result is different from this}

Overall, with our control perturbations based co-location, we successfully co-located 10 out of 16 point types (63\% coverage) with 98.4\% accuracy across eight VAV units. We co-located the \emph{Occupied Command} points using auxiliary control experiments for four zones. We failed to co-locate \emph{Damper Command} due to its divergent behavior. We also do not co-locate 4 point types which do not respond to control perturbations. %Thus, our coverage of co-location for VAV points is 63\%. \YA{change this to talk about coverage per zone first and then say that the 100\% accuracy number is for colocating the same 10 points across all zones in the building.}
%\BB{Add relaxation of point type assumption if we have the analysis}

\subsection{Identification of Point Type}

We now look at the inverse problem of identifying point type given the co-located points in a VAV. Identification of point type is essential for third party application to interpret timeseries data, and prior work has shown that data analysis alone fails to identify point types accurately~\cite{gao2015metadata,hong2015adapter}.

As shown towards end of Section \ref{sec:colocation}, it is possible to identify co-located points even when the individual point types are unknown. For this problem, we assume that we know the co-location of all the VAV point types and the ground truth point types for one zone. We use this information along with the timeseries data to identify point types of other zones in the building. When our data analysis fails to identify certain point types, we use control experiments to increase the coverage of point types identification. Note that the VAV points we focus on cover 79\% of all points in our building testbed.

\begin{figure}[Ht]
\includegraphics[width=\linewidth]{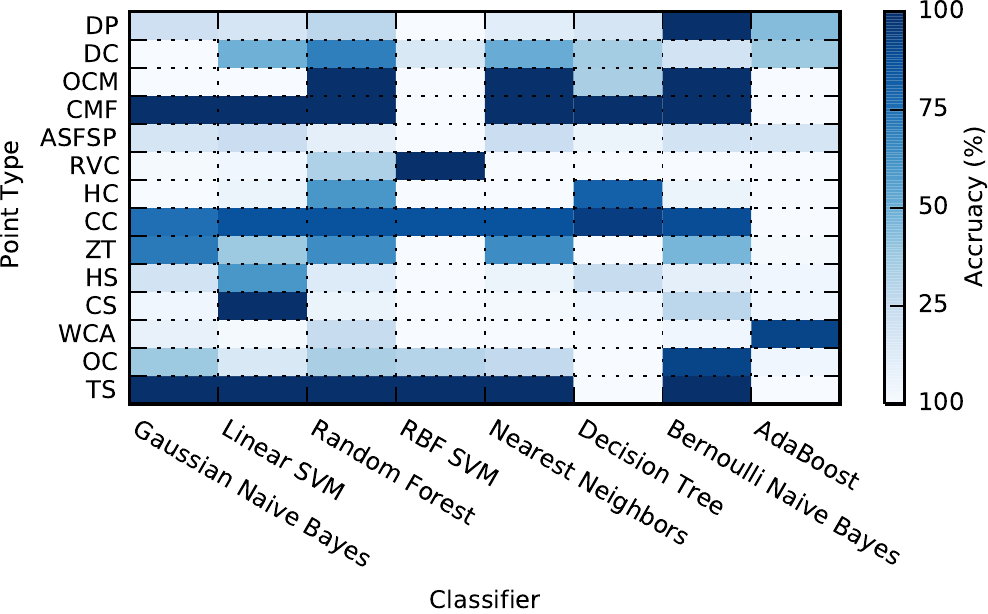}
%\vspace{-5mm}
\caption{Accuracy of point type identification using transfer learning across ten supervised classifiers with one year historical data for VAV points belonging to one zone.}
\label{fig:type_data}
%\vspace{-5mm}
\end{figure}

With the help of the zone for which ground truth is known apriori, we train supervised classifiers using features extracted from one year of timeseries data and ground truth point types as labels. We slice one year of data into 53 weeks (partial weeks for starting and ending week), and extract features such as mean, variance, dominant frequency, noise, skewness and kurtosis~\cite{tsay2005analysis} for each week. Noise is represented by error variance between original data and its piece-wise contant approximation. Skewness measures the symmetry of the data values across the mean while kurtosis measures the peak of the data distribution at the mean. The vector of features for one week of data represents one row of training data for our classifier. We train many standard classifiers -- Gaussian Naive Bayes (GNB), Linear Support Vector Machine (LSVM), Random Forest (RF), Radial Basis Funcation SVM (RBF SVM), Nearest Neighbours (NN), Decision Tree (DT), Adaboost, and Bernoulli Naive Bayes (BNB). Our test results show high accuracy ($\sim$100\%). We use these trained models for identifying point types of \emph{different} zones with the same features extracted from one year of data. This process is called \emph{transfer learning}, and figure \ref{fig:type_data} summarizes the results obtained across all the classifiers for different point types in one zone. 

We find that some point types such as \emph{Temperature Setpoint} (TS) are readily identified even with simple features such as mean and variance across most classifiers, while other point types work only with specific classifiers or require more features. We experimented with 10 classifiers for three zones, and choose the classifier which works best for each point types for analysis with other zones. Some point types such as \emph{Heating Command} (HC), \emph{Cooling Command} (CC) and \emph{Zone Temperature} (ZT) are not identifiable by any classifier with all the features we used. This is because the behavior of these point types are similar to each other or other identified point types during regular operation. We repeated this analysis across 8 zones and found similar results. Our results are summarized in Table \ref{tab:type_results}. These results are corroborated by prior work who used more sophisticated data features for type identification~\cite{balaji2015zodiac,gao2015metadata,hong2015adapter}.

\begin{table}[!t]
\renewcommand{\arraystretch}{1.3}
% if using array.sty, it might be a good idea to tweak the value of
% \extrarowheight as needed to properly center the text within the cells
\caption{Results of point type identification using both data analysis and analysis with control perturbations.}
\label{tab:type_results}
\centering
% Some packages, such as MDW tools, offer better commands for making tables
% than the plain LaTeX2e tabular which is used here.
\begin{tabular}{|c|c|c|c|c|}
\hline
\textbf{Point Type} & \multicolumn{2}{c|}{\textbf{Data}}  & \multicolumn{2}{c|}{\textbf{Control}}\\
\hline
 & Classifier & Accuracy & Classifier & Accuracy\\
\hline
Temperature Setpoint & RF & 100\% & - & - \\
\hline
Occupied Command & RBF SVM & 99.2\% & - & - \\
\hline
Thermostat Adjust & AdaBoost & 93.1\% & - & - \\
\hline
Cooling Setpoint & LSVM & 88.9\% & - & - \\
\hline
Heating Setpoint & DT &87.4\% & - & - \\
\hline
Zone Temperature & RF & 65.0\% & RF & 88\% \\
\hline
Min Supply Flow & RF & 77.8\% & - & - \\
\hline
Max Supply Flow & BNB & 100\% & - & - \\
\hline
Cooling Command & DT & 47.0\% & DT & 88\% \\
\hline
Heating Command & DT & 32.1\% & - & - \\
\hline
Reheat Valve Cmd & RBF SVM & 78.4\% & - & - \\
\hline
Supply Air Flow SP & NN & 80.3\% & - & - \\
\hline
Supply Air Flow & LSVM & 90.6\% & - & - \\
\hline
Damper Command & RF & 96.2\% & - & - \\
\hline
Damper Position & RF & 80.3\% & - & - \\
\hline
\end{tabular}
\end{table}

We leverage control perturbations to identify point types which were unsuccessful using data analysis. As points such as TS and \emph{Occupied Command} (OC) are easily identifiable using data analysis, we use these points to create our control perturbations. We put the zone to the \emph{Occupied} mode, and increase the TS to $78^oF$ to force the VAV into a heating mode for 3 hours. The same perturbation is applied for the ground truth zone as well as all the zones for which the point type needs to be identified. We then extract the same data features for this control period across all the points, and use transfer learning to label points in different zones. As the behavior of all the controlled zones is forced to be similar, the behavior of the same point type across the zones is similar. However, the controlled sequence also forces the non-identified point types -- SAF, SAFS, ZT, HC -- to be different from the rest of the points. Thus, with this experiment, we are able to identify the rest of the point types successfully. Our results across the 8 zones have been summarized in Table \ref{tab:type_results}. Overall, our accuracy of point type of identification is 85.3 \%.

Our data analysis or control experiments could not distinguish between point types \emph{Heating Command} (HC) and \emph{Reheat Valve Command} (RVC) as they are virtually identical in their behavior. However, it is possible to separate these two points using dependency analysis, which we describe next.

\subsection{Dependency Mapping of Points}
\label{sec:dependency}

We now focus on the understanding the working of the VAV, and how the points relate to each other. As the type of points exposed are different across vendors and equipment, it is necessary to understand the context of these points, and map it to a model that can be used by other applications. These models can be built using domain knowledge, technical documents and historical data analysis~\cite{foucquier2013state,jensen1996introduction} as demonstrated by our dependency graph in Figure \ref{fig:dependency_vav}. We propose control perturbations as an alternative to these methods, which can be used for either verifying already developed models or used for older buildings where information available is insufficient for modeling using other methods.

We assume that we already know the point type and the co-located points in a VAV. We focus on modeling the dependencies between the actuator points (or read/write points). %\JK{Don't we need to clarify that we keep changing of TS to track ZT? I tried to put some explanation about that, but it is quite unclear to say it. Need discussion of this.} Under our framework, 
%We exploit our control framework's benefit as it can guarantee that there is no external perturbation of the system other than internal system control except we keep ZT to be between ACS and AHS not to cause other control. This is required but minimal knowledge, but it is obvious that a VAV controls itself to set a ZT to TS. Then, we can say that our control inputs are the root of the causal chain. \JK{This already contains our knowledge and much assumption...}

We write to the \emph{Temperature Setpoint} (TS) of a zone with a randomly chosen value every 20 minutes for 6 hours. The goal of this control sequence is to identify the points that react to changes in TS, and so we choose random values within limits for perturbing different operating points of the control system. For every change in TS, we analyze the behavior of the VAV points for 10 minutes, and note the points whose values change during this period. The threshold of change is one standard deviation for values observed for the past 12 minutes. We chose these times so that we can isolate changes that occur due to the change in TS rather than other external factors such as solar radiation. For the duration of the experiment, we calculate a final probability for each of the points as the ratio of the number of changes observed for the point and the number of changes in TS. We repeat this experiment by perturbing all the actuator points - \emph{Occupied Command} (OC), \emph{Cooling Command} (CC), \emph{Heating Command} (HC), \emph{Supply Air Flow Setpoint} (SAFS) and \emph{Damper Command} (DC). Note that we ignore the points related to minimum and maximum supply air flow as they are constants. %\YA{again, where are these numbers 20min, 6 hours, etc from?} 

\begin{figure}[Ht]
\includegraphics[width=\linewidth]{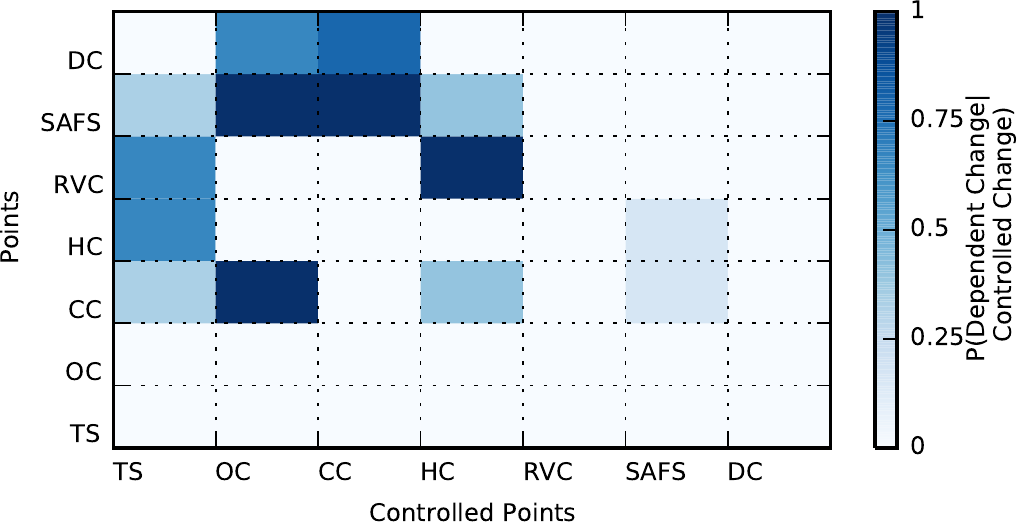}
%\vspace{-5mm}
\caption{Color map showing the changes induced by control perturbations of each actuator point on other actuators. Probabilities are calculated as the ratio of number of changes observed in a non-controlled actuator and the number of changes made by the controlled actuator.}
\label{fig:dependency_cmap}
%\vspace{-5mm}
\end{figure}

Figure \ref{fig:dependency_cmap} shows a color map representing the probabilities obtained by perturbing each of these points. The changes to TS affects all the actuators except OC and DC, while changes to actuators like CC cause changes only in DC and SAFS. With the help of this color map, we can understand which points are being affected by each of the actuator points. However, this does not precisely decide the dependency between points as points which are lower in the dependency tree such as SAFS get affected by almost all of the actuation experiments. We find the behavior of DC to be unpredictable, and the changes that occurred in DC with our control perturbances were lesser than our set threshold. We perform these perturbation across five zones.%We focus on exploiting data analysis based dependency analysis to augment our methods in future work.

\begin{figure}[Ht]
\includegraphics[width=\linewidth]{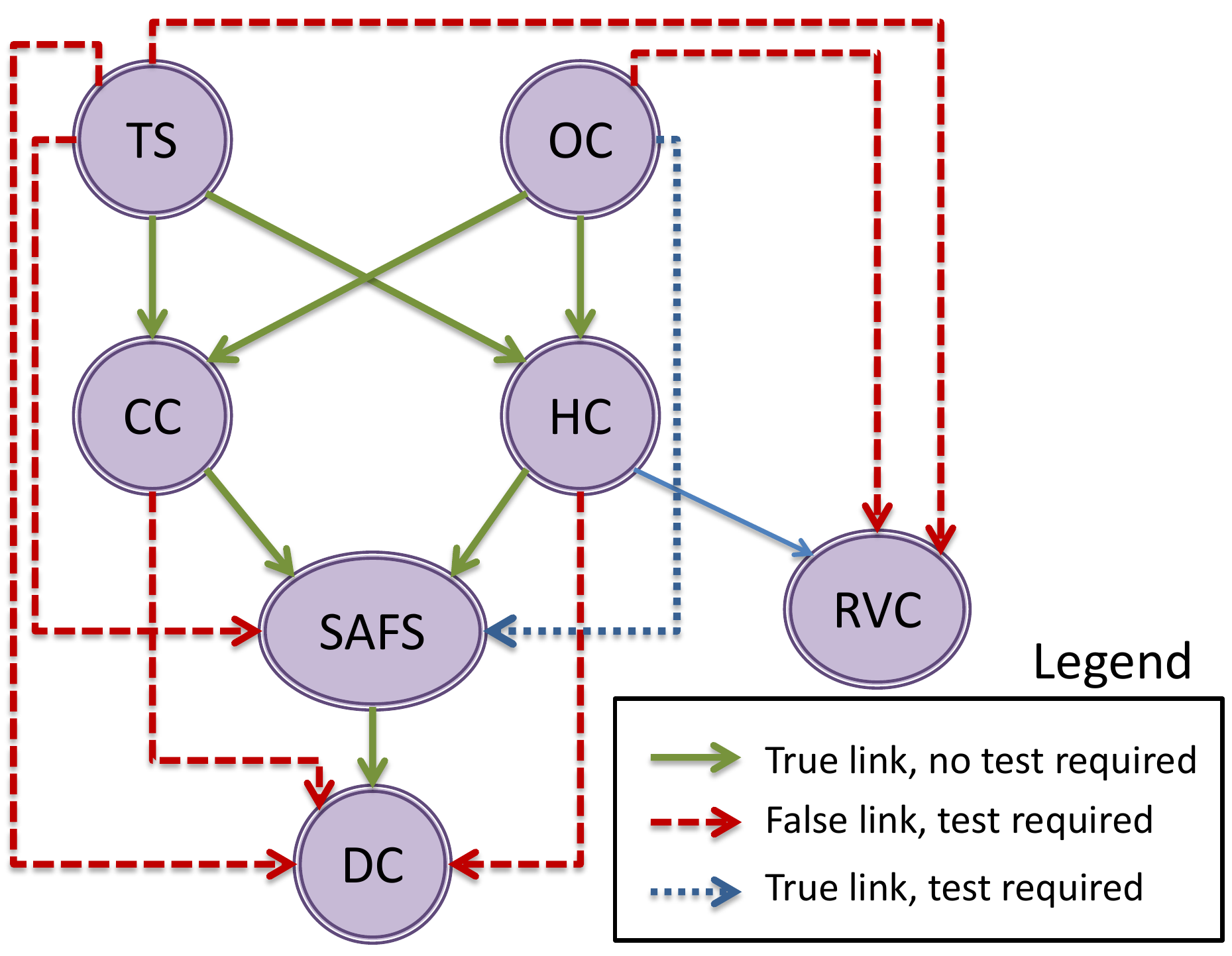}
%\vspace{-5mm}
\caption{Dependency links obtained by perturbing each of the actuators and checking which other actuators react to these changes. The links which require further testing are verified using conditional control perturbations.}
\label{fig:dependency_links}
%\vspace{-5mm}
\end{figure}

Figure \ref{fig:dependency_links} shows the relationships obtained as a result of our analysis. The green solid links indicate relationships which are true and confirmed with the analysis. The red dashed links show relationships which are not true, but are shown to be related by the analysis. In general, the above experiments cannot identify the true links when a ``cycle'' is formed in the graph. Here, by ``cycle'' we mean that there are multiple paths from one point to another in the graph. We perform more control experiments to negate the red links, and also confirm the blue dotted links which form a cycle with the green links. %\YA{how is the dotted blue forming a cycle with the green?} 

Consider the cycle formed between the points TS, CC and SAFS. In order to verify if the link between TS and SAFS is correct, we perform a conditional control perturbation, where we change TS but force CC to be unchanged, which is called \emph{Graph Surgery}~\cite{pearl2009causality}. We repeat this experiment for at least four changes of TS. If TS were directly affecting SAFS, we would observe that SAFS changes even when CC is held constant. We verify each of the red and blue links this way. When there are more than three points in a cycle, such as that with TS, OC, HC and RVC, we ensure several combinations of TS and OC are performed to test validity of the link. In some of these experiments, we preset the TS value to a fixed value for appropriate conditions that can activate other points such as HC. As we note at Section \ref{sec:framework}, an external variable, ZT, may disable HC or CC in certain condition though ZT is not an actuator. We let the VAV control system to settle to a steady state after our change of TS before performing any dependency experiments.

We performed these experiments on five zones in our building, and verified the links with 73.5 \% accuracy with a false positive rate of 18.4 \% and false negative rate of 8.1 \%. All of the false positive and  64.7 \% of the false negative are due to the external variable, ZT. 
Thus, we can discover the dependency between actuator points in the VAV using our control experiments. However, in order to discover the complete dependency map as shown in Figure \ref{fig:dependency_vav}, we need to use maximum likelihood data analysis as the behavior of read-only points cannot be controlled directly. %We can model these dependencies by using maximum likelihood analysis on the historical data. These can be augmented with control perturbations for observability of unknown behavior. We pursue these techniques in future work. 

%\subsection{Relationship Between Points}

%\subsection{Anomaly Detection}
\section{Related Work}
\label{sec:related}
The problem of discovering system characteristics and models using available data is called \emph{system identification}~\cite{ljung1998system}, and is a well studied subject in control systems research. Using control perturbations for system identifications is also well known~\cite{ljung1998system,godfrey1993perturbation}, and the design of control perturbations, also referred to as auxiliary or secondary signals, has been studied for modeling different types of control systems~\cite{godfrey1999comparison,godfrey2005survey}. System identification techniques have also been used for HVAC modeling~\cite{teeter1998application}, and some prior work have explored using control signals for system identification~\cite{sousa1997fuzzy,virk1995practical}. However, all of these works focus on modeling the control system or perform control optimizations and do not address identification of contextual information such as location, point type or dependency graphs. Moreover, the control perturbation methods used for buildings are only verified using simulations.

Active HVAC control on real systems has been used for fault diagnosis~\cite{weimer2012active} and fault tolerant control~\cite{fernandez2009self}. These works and other simulation based studies which propose fault tolerant control~\cite{liu2001fault,wang2002fault} assume the contextual information about the system is already available. We focus on discovering contextual information using active control. 

Co-location of sensors has been studied before~\cite{fontugne2012empirical,hong2013towards} but they use sophisticated data analysis algorithms. These methods fail when the points from different locations have similar data characteristics during regular usage. We show that with perturbations we can excite the local control system to unique operating points and co-locate points with high accuracy using simple data analytics. 

Point type identification has also been studied earlier using both metadata~\cite{balaji2015zodiac,bhattacharya2015metadata} and data analytics~\cite{gao2015metadata,hong2015adapter}. These works show that metadata alone can be unreliable and requires significant manual input for accurate type identification, and the data analytics based method is useful for some point types but fails for others. Our results conform to the data analysis works, and we show that perturbations can be used to identify the points which are difficult with data analysis. We also note that Hong et al.~\cite{hong2015adapter} focus on transfer learning across buildings, while we focus on transfer learning within the building across the different instances of VAV units.

In addition, we focus on creating dependency graphs between points in the HVAC system. Pritoni et al.~\cite{pritoni2015discovering} use control of AHU discharge air temperature to map the VAV units to their corresponding AHUs. We present control techniques to isolate relationships within the VAV unit. %In addition, we use control for co-location and point type identification for VAV points.

We have presented a detailed view of how HVAC VAV units work in practice, and how our Quiver control framework has been designed for safe control of the VAV unit, and provides features such as query of current status and rollbacks in case of crashes. Dawson-Haggerty et al.~\cite{dawson2013boss} provide similar mechanisms to provide safe operations, but the onus of rollbacks and error checking is on the application developer. They also provide support for multi-user control with timed leases and locks, while we only support single user control. 

%- System identification and control perturbations

%- Past CPS work which focused on control. What types of work, and how they actually use control.

%- Past buildings work that focused on control. Berkeley paper that exploited control~\cite{pritoni2015discovering}.

%- Active fault diagnosis, automated functional testing and fault tolerant control
\section{Limitations and Future Work}

We have shown that control perturbations can be useful in discovery of contextual information in HVAC VAV units in our building testbed. This is just a first step, and more research is needed to extend these results to other types of equipment and control systems. We hope that the research community embraces control perturbations as a tool and generalizes the results across different buildings, vendors and even other cyber physical systems.

Although our control perturbations were simple and effective for extracting information we were interested in, we observe that design of perturbation signatures can affect the results significantly depending on external disturbances and behavior of the control system. We will focus on automating the perturbance signals in future work, borrowing ideas from system identification literature~\cite{godfrey2005survey}. We have also forced our control experiments to be done on either weekends or nights. Our work needs to be extended to perform these experiments even during work hours with perturbation signals that conform to comfort constraints. 

To fully exploit control capability for both perturbation applications and control optimizations, we need to provide support to third party developers for direct control of systems. Initial work in both academia~\cite{dawson2013boss} and industry~\cite{niagara} have extended traditional vendor specific controls to common APIs. Our control framework, Quiver, allows safe control by a single user. Further work needs to be done to provide safety guarantees so that multiple users can access the control system at the same time and developers are provided sandboxes or emulation tools to experiment with different control applications. 

%- Expand Quiver to include AHU and other types of equipment

%- Allow Quiver to be used by multiple developers at the same time. Providing protection and transaction guarantees.

%- Extending the analysis to more system identification and auxiliary signal analysis. Control perturbations can be automatically constructed.
\section{Conclusion}

Traditional Building Management Systems are vendor specific vertically integrated solutions. To enable third party building applications that exploit sensor information, recent works have created standardized APIs and data management solutions which have led to a spurt of innovative applications. We extend these works by enabling control of building systems and demonstrate applications that exploit control to extract contextual information. We design Quiver, a control framework that allows safe experimentation of HVAC VAV units by integrating domain knowledge gained through experience and empirical experiments. Using Quiver, we demonstrate three control perturbation applications that extract context information about VAV units -- co-location of points, identification of point types and mapping the dependency between points. We co-locate 63\% VAV points with 98.4\% accuracy, we identify point types with 85.3 \% accuracy across 8 zones, and we identify dependencies between VAV actuator points with 73.5 \% accuracy across 5 zones. 

\bibliographystyle{IEEEtran}
\bibliography{IPSN2015-Quiver_arXiv}
%\begin{thebibliography}{1}
%\end{thebibliography}

\end{document}